\documentclass[twocolumn]{webofc}

\usepackage[varg]{txfonts}   
\usepackage{hyperref}
\usepackage{url}
\usepackage{gensymb}
\usepackage{ragged2e}

\usepackage{acronym}
\acrodef{HCAL}{Hadronic Calorimeter}
\acrodef{DHCAL}{Digital Hadronic Calorimeter}
\acrodef{RPWELL}{Resistive Plate WELL}
\acrodef{GNN}{Graph Neural Network}
\acrodef{GAT}{Graph Attention Transformer}
\acrodef{PFA}{Particle Flow Algorithm}
\acrodef{ML}{Machine Learning}
\acrodef{MLP}{multilayer perceptron}

\hypersetup{colorlinks=true,citecolor=blue,urlcolor=blue,linkcolor=blue}
\begin{document}
\title{Point Cloud Deep Learning Methods for Particle Shower Reconstruction in the DHCAL}

\author{\firstname{Maryna} \lastname{Borysova}\thanks{\email{maryna.borysova@weizmann.ac.il}}, \
        \firstname{Shikma} \lastname{Bressler},\
        \firstname{Eilam} \lastname{Gross},\
        \firstname{Nilotpal} \lastname{Kakati}\ and
        \firstname{Darina} \lastname{Zavazieva}\
}

\institute{Weizmann Institute of Science, Rehovot, Israel}

\abstract{Precision measurement of hadronic final states presents complex experimental challenges. The study explores the concept of a gaseous \ac{DHCAL} and discusses the potential benefits of employing \ac{GNN} methods for future collider experiments. In particular, we use \ac{GNN} to describe calorimeter clusters as point clouds or a collection of data points representing a three-dimensional object in space. Combined with \acp{GAT} and DeepSets algorithms, this results in an improvement over existing baseline techniques for particle identification and energy resolution.
We discuss the challenges encountered in implementing \ac{GNN} methods for energy measurement in digital calorimeters, e.g., the large variety of hadronic shower shapes and the hyper-parameter optimization. We also discuss the dependency of the measured performance on the angle of the incoming particle and on the detector granularity. Finally, we highlight potential future directions and applications of these techniques.
}

\maketitle
\section{Introduction}
\label{intro}

In particle physics, accurate measurement of jets is crucial for understanding fundamental properties of particles and their interactions. Precision measurement of hadronic showers poses significant challenges.
The most prominent one originates from the different response of most \ac{HCAL} to hadrons and electrons, i.e., e/h ratio different from 1~\cite{Wigmans}. This complicates the task of obtaining a direct measurement of the jets energy based solely on the signals detected. This problem is enhanced due to the large event-by-event fluctuations of hadronic showers attributed to large variations in the fraction of invisible energy, the energy transfer from hadronic to electromagnetic component via $\pi_{0}$ production, and other phenomena. 

A promising approach to calorimetric measurement of jet energies is the \ac{PFA} \cite{Thomson:2009rp}. Using small calorimeter cells (large granularity) enables to track charged particles in the calorimeter and associate their energy deposits with the corresponding tracks from the tracking system, which measures their momenta with high precision. After removal of this “charged-particle energy" from the calorimeter information, the remaining energy is measured using the residual hits in the calorimeter, namely those without associated incoming charged-particle tracks. Thus, an experiment optimized for the particle flow approach relies on the \ac{HCAL} mostly for the measurement of the energy of neutral hadrons (mostly neutrons) – about 10-20\% of the total jet energy – and the deterioration of the jet energy resolution due to the poor intrinsic resolution of the \ac{HCAL} is mitigated. 

Sampling \ac{HCAL} with digital (1 bit) readout, i.e., \ac{DHCAL}, consists of alternating layers of absorbing material and sensitive elements where the hadronic shower is formed and measured, respectively. For each shower, the output of such calorimeters is a set of fired pads and their position. Traditional energy measurement algorithms convert the number of measured hits to the energy of the incoming particle~\cite{Dans, Fe_calice}. More advanced algorithms take into account also the density of hits in the different layers. A promising approach to improve the energy measurement even further is utilising \ac{ML} techniques, within which we focus on employing \ac{GNN}.

\begin{figure*}
\centering
\vspace*{1cm}       
\includegraphics[width=10cm,clip]{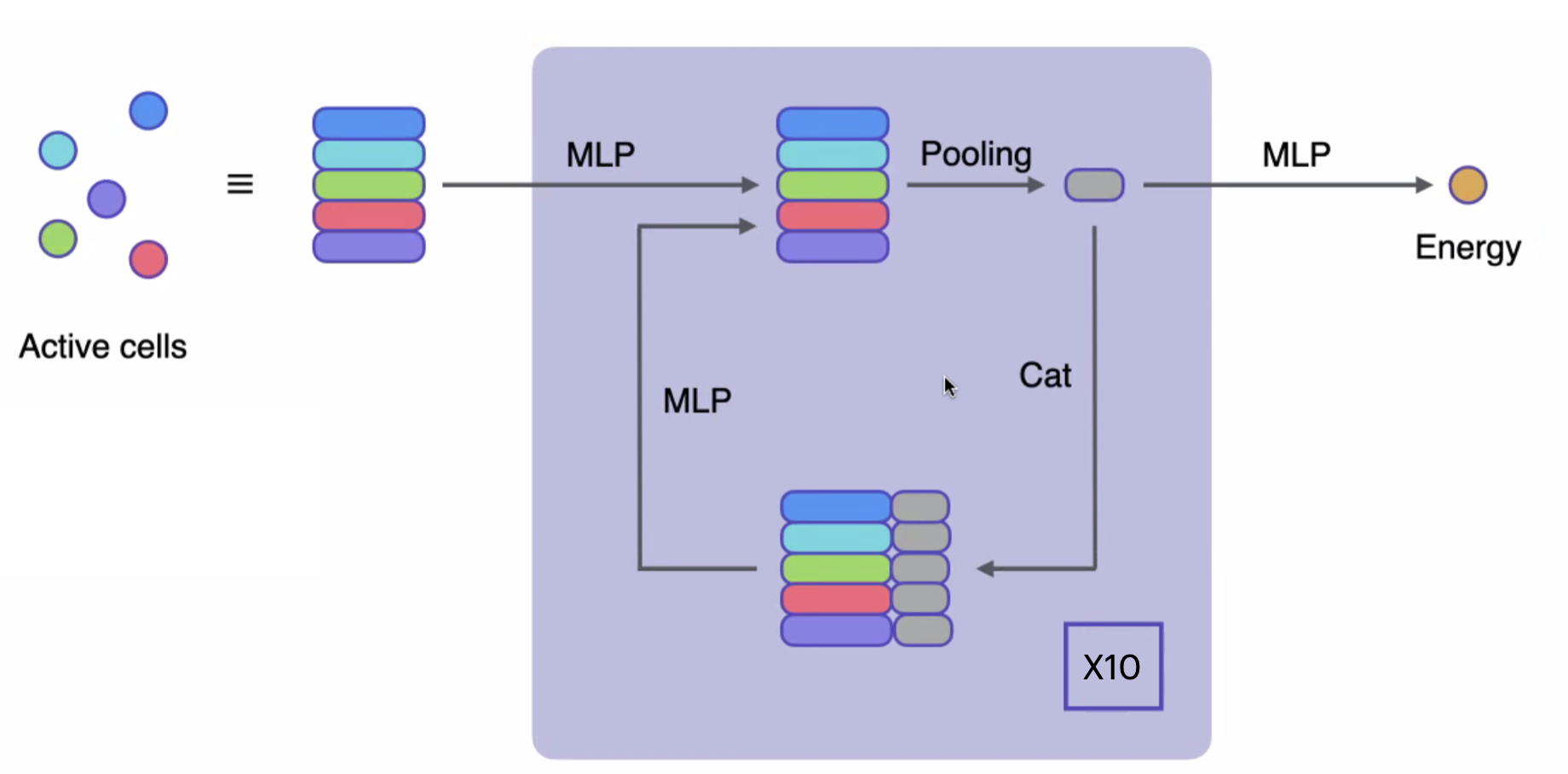}
\caption{DeepSets architecture. The activated calorimeter cells, coloured circles, represent the point clouds. Each point cloud is independently processed by a \ac{MLP} to extract the vector of features (coloured ovals) - the node position in space. The feature vectors from all cells are aggregated into a single representation using average Pooling (grey oval). Concatenation (Cat) is used to combine different feature vectors derived from the input set into a single vector that represents the entire set. The aggregated representation is further processed by an \ac{MLP} to generate the final output, such as the energy.}
\label{fig:deepset}   
\end{figure*}

\section*{Methods}
\label{methods}
1.2M hadronic showers of different particle types (charged pions, long-lived neutral kaons, protons and neutrons)  with energies from 1 to 60 GeV  were simulated using a custom GEANT4\footnote{version 10.06.p01  \cite{geant4}, with QGSP-BERT-EMZ physics list} model of a fully-equipped (50 layers) \ac{RPWELL}-based \ac{DHCAL} with 2-cm-thick steel absorbers and $1 \times 1~cm^2$ pad readout \cite{Dans}. These 50 layers correspond to a total depth of $~5\lambda_{\pi}$, inferring a 99.3\% chance for a pion to initiate a shower within the calorimeter, thus ensuring minimal energy leakage. 
The expected performance of the \ac{DHCAL} was evaluated based on past measurements with smaller \ac{RPWELL} prototypes \cite{Dans} which showed MIP detection efficiency of 98\% and pad multiplicity of 1.1. To minimize the longitudinal leakage of the shower, the events are pre-selected for the analysis with the identified shower start in the first 10 layers of the calorimeter.
After pre-selection for each particle type, the data set contains ~600k events that were used for the NN training, 100k events for validation and another set of 110k events, not seen by the network, for shower shapes testing.

We explored two NN architectures in this study: DeepSets ~\cite{zaheer2018deepsets} and \ac{GAT} ~\cite{vaswani2023attentionneed}. Figure \ref{fig:deepset} illustrates the DeepSets architecture schematically. 
It is designed to handle sets of individual data points (nodes), often referred to as a point cloud. In the context of calorimetry, this point cloud represents the activated calorimeter cells, each signifying a location where a particle has interacted with the detector material.

Unlike DeepSets, which process point clouds by focusing solely on nodes, the GAT approach incorporates both nodes and edges (relationships between data points) to leverage relational information. This allows the GAT layers to exploit the inherent structural information within the shower data. We employ a masked attention mechanism. This restricts information sharing between nodes to only geometrically close neighbours (within a cone of radius 0.1). This allows the model to focus on relevant local information while still considering global context.
In both architectures, the input cells are first processed by a \ac{MLP} to extract relevant features. These features are then refined through multiple iterations of either DeepSets or GAT layers. Finally, the refined cell representations are aggregated into a global representation, which is used to make the final prediction, such as the energy of the incident particle or its classification.
 
The NNs  were implemented in PyTorch~\cite{pytorch} and Adam optimizer was used to improve the convergence rate.
To optimize the training process, several hyper-parameters~\cite{tensorflow}, such as the learning rate and batch size, were explored. The learning rate controls the step size taken during gradient descent, while the batch size determines the number of training samples processed before updating the model's parameters. 

While full hyper-parameter optimization can enhance model performance, the parameters used along this project e.g., a learning rate of $10^{-4}$ was found to yield results sufficiently good to demonstrate the strength of this approach. Different batch sizes were also tested in the range from 1 to 256 events per batch. No instabilities were observed as the batch size increased. The batch size of 64 has been chosen as a compromise between the convergence stability and the speed of calculation.
Two target variables are investigated: energy and particle type. Energy prediction is treated as a regression task and optimized using mean squared error (MSE) loss. For particle identification, a multi-class classification approach is employed. The model outputs a probability distribution over $n$ possible particle types, where $n$ is the number of potential incident particle candidates. The probabilities are obtained by applying the Softmax function 4.1~\cite{softmax} to the model's output. The particle type with the highest probability is selected as the final prediction.

\section*{Results}
\label{results}
\begin{figure}[h]
\centering
\includegraphics[width=8cm,clip]{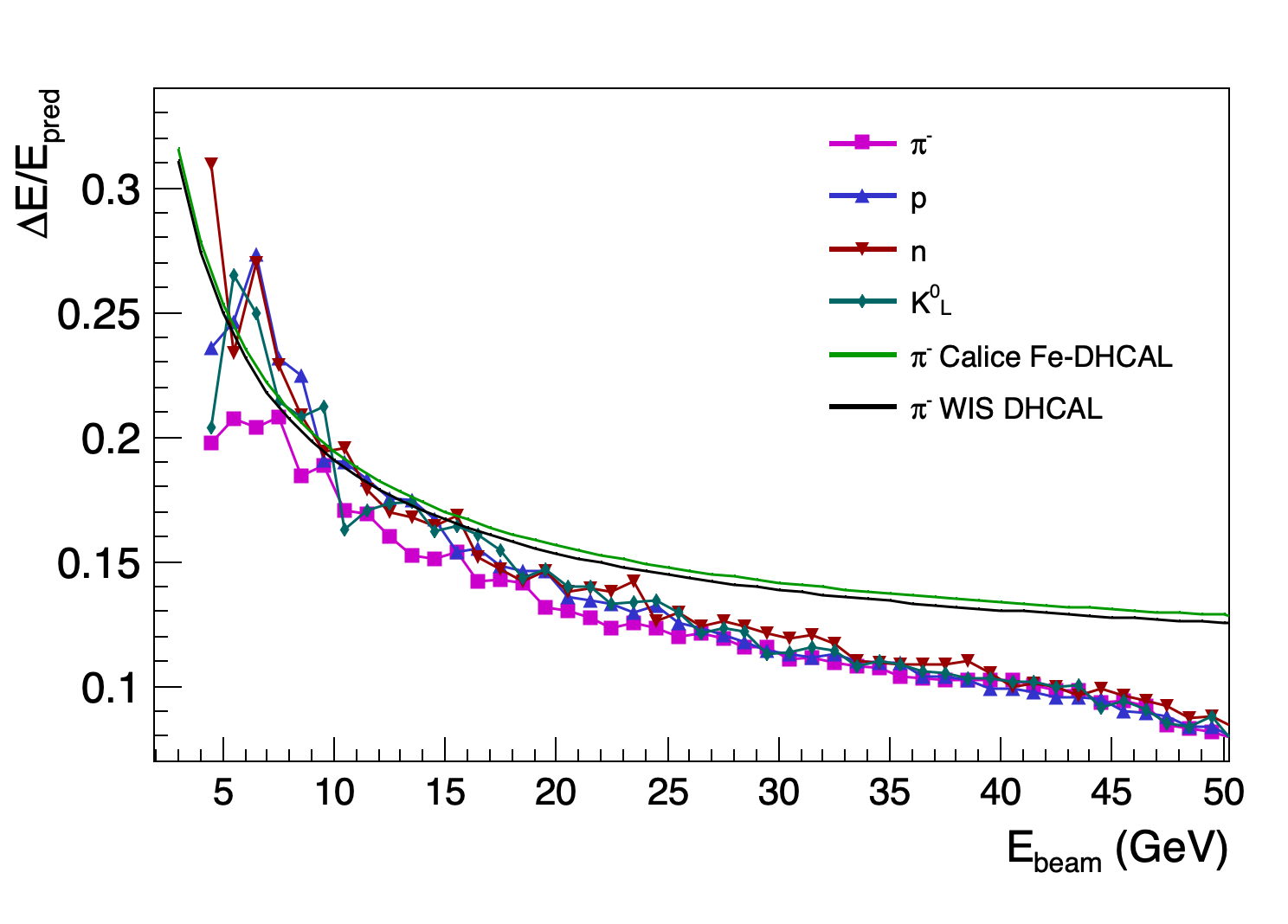}
\caption{Predicted energy resolution for various incident hadron types as determined by the DeepSets architecture.}
\label{fig:res_particle}
\end{figure}

In this section, we discuss the application of \acp{GNN} for energy reconstruction and particle identification, utilizing the two NN architectures.
The energy resolution obtained with DeepSets architecture for the different hadron types traversing perpendicular to the calorimeter (e.g., 0 degree angle) is shown in Figure~\ref{fig:res_particle}. The energy resolution of pions (magenta line) outperforms traditional algorithms for the \ac{RPWELL}-based \ac{DHCAL} \cite{Dans} (black line) and RPC-based CALICE \ac{DHCAL} \cite{Fe_calice} (green line) data. The energy resolution curves of neutrons, kaons and protons are also shown.

\begin{figure}[!h]
\centering
\includegraphics[width=7.8cm,clip]{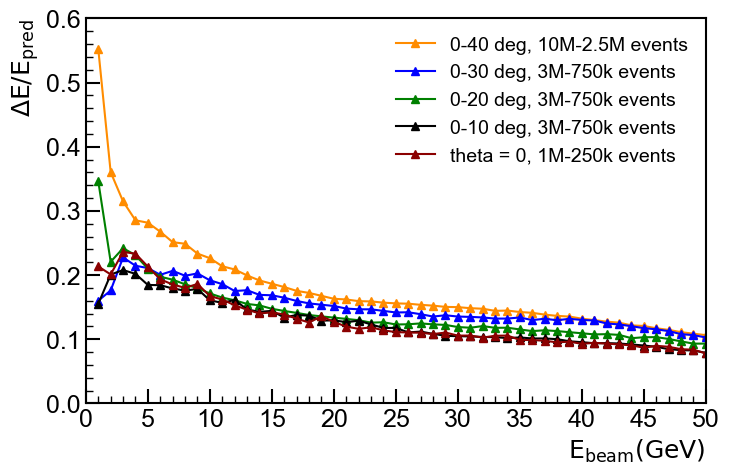}
\caption{Energy resolution predicted by DeepSets for negative pions entering calorimeter at various incident angles.}
\label{fig:res_angles}
\end{figure}

The shape and a shower development vary with the incident angle. This affects the energy deposition patterns in the \ac{DHCAL} and thus, the energy resolution. We studied this dependency by training the NNs with pions uniformly distributed in various angle ranges. The results are depicted in Figure~\ref{fig:res_angles}. For the tested NNs, the energy resolution of pions traversing within a cone larger that 20\degree~deteriorates. This could be attributed to lateral leakage and to the need of bigger training dataset when considering larger variability of shower shapes. 

\begin{figure}[!h]
\centering
\includegraphics[width=8cm,clip]{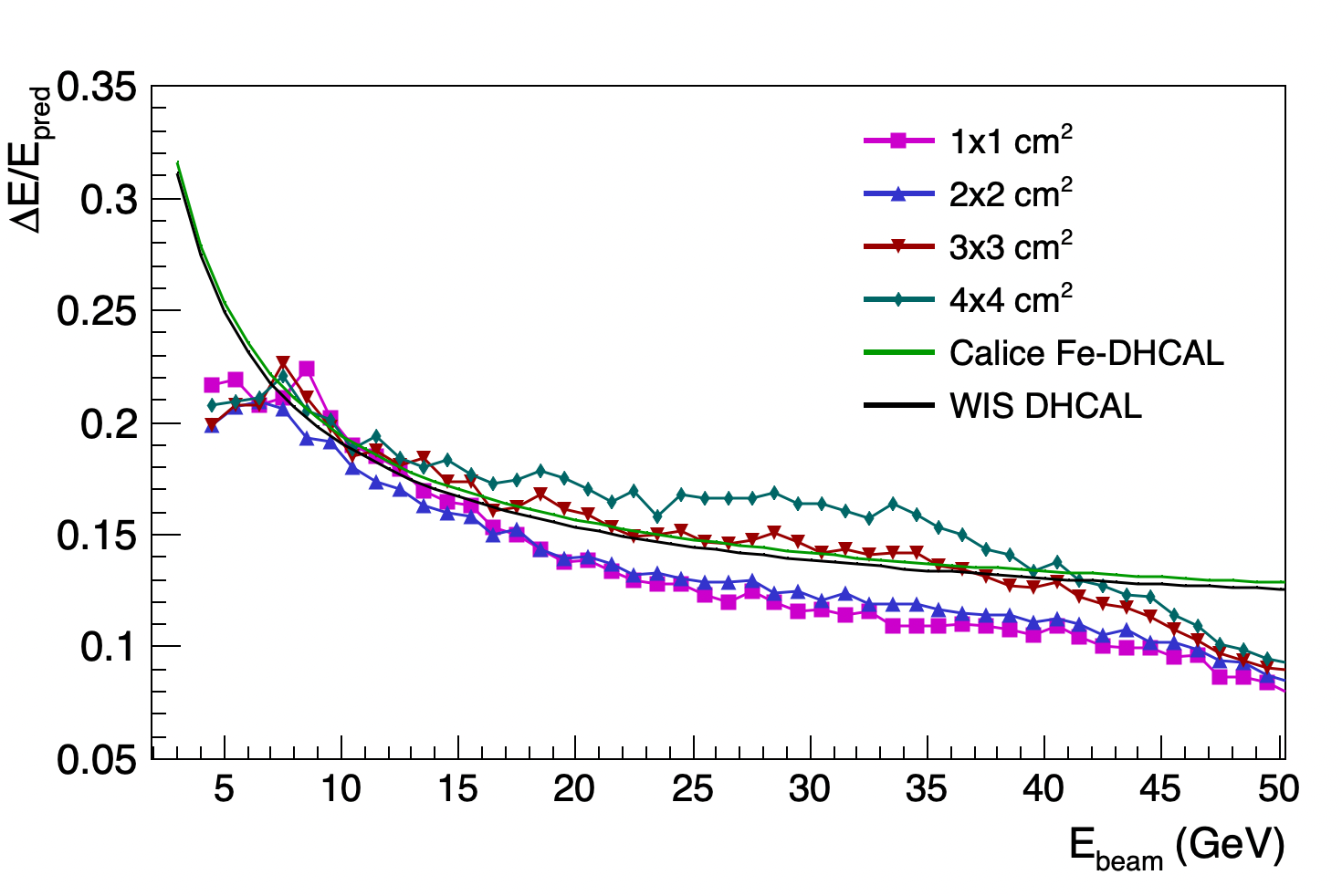}
\caption{Predicted energy resolution for pion showers using the DeepSets model as a function of the \ac{DHCAL} pad size. The results demonstrate the model's sensitivity to the granularity of the detector.}
\label{fig:res_pad}
\end{figure}

The dependency of the pion energy resolution for various pad sizes is shown in Figure~\ref{fig:res_pad}. While the baseline design of such calorimeter is pad size of $1 \times 1~\mathrm{cm^2}$, enlarging the pads by a factor of four ($2 \times 2~cm^2$) and reducing the number of channels by four does not degrade the performance significantly. Provided that the two shower separation would not degrade as well, these may offer more cost-effective solution for future experiments.
GAT achieves similar performance to DeepSets, but at a significantly higher computational cost (3 times slower) and GPU memory usage (10 times more).

\begin{table}[!h]
    \centering
    \begin{tabular}{c|c|c|c|c|}
         & \multicolumn{4}{c|}{Predicted} \\ \hline
         True  & Neutrons & Pions  &  Protons  & Kaons\\ \hline
         Neutrons&  0.55  & 0.02 & 0.05 & 0.39 \\
         Pions   &  0.06  & 0.51 & 0.34 & 0.10 \\
         Protons &  0.10  & 0.26 & 0.57 & 0.07 \\
         Kaons   &  0.30  & 0.02 & 0.02 & 0.65 \\
    \end{tabular}
    \caption{\raggedright Table illustrating the performance of the GAT model in identifying different particle types. The diagonal elements represent the correctly classified particles (true positives), while off-diagonal elements indicate misclassifications.}
    \label{tab:id}
\end{table}
While both DeepSets and GAT architectures demonstrate promising results in energy reconstruction, their performance in particle identification differs. DeepSets exhibits limitations in accurately identifying particle types, a challenge we aim to address in future studies. In contrast, GAT shows significant potential for PID, as demonstrated by its ability to distinguish between neutrons, pions, protons, and kaons. As shown in Table~\ref{tab:id}, the model assigns higher probabilities to the correct particle type, particularly for protons and kaons. 
This is due to the fundamental differences in the interaction mechanisms of studied hadrons with the detector material. 
The difference is attributed to baryon number conservation in proton showers and strangeness number conservation in kaon induced showers, which reduces event-to-event fluctuations in the production of of neutral pions $\pi_0$ particles — key carriers of energy in electromagnetic components of hadronic showers.
The conservation of baryon  number and strangeness restricts the possible decay and interaction channels, leading to a more controlled production of secondary particles and predictable energy deposition patterns in a calorimeter.
The lower score for neutrons could be due to the fact that neutrons can undergo charge exchange reactions, where they interact with protons in the detector material to produce charged pions. This can lead to variations in shower shapes and energy deposition patterns.

\section*{Summary}
\label{summary}
In this work, we explored the potential of \ac{GNN} to improve the design and performance of sampling \ac{DHCAL} with 1 bit-readout. We used a GEANT4 model to simulate the response of a fully equipped \ac{RPWELL}-based \ac{DHCAL} to pions, protons, kaons, and neutrons at a variety of energies, angles and pad readout sizes. We discuss the application of \acp{GNN} to energy reconstruction and  particle identification in the framework of two NN architectures, DeepSets and Graph Attention Transformers. In the DeepSets architecture, the fired pads are treated as individual data points, and their spatial relationships are not explicitly modelled. In contrast, GATs leverage the spatial relationships between pads by representing them as nodes in a graph, where edges connect neighbouring cells within a defined radius.
We have shown that the energy resolution does not degrade with four times larger readout pad-size and corresponding and channel reduction which could lead to much more cost-effective solutions for future sampling calorimeters.

The results demonstrate the potential of deep learning techniques for hadronic showers energy reconstruction and particle identification and are therefore an important step towards the optimisation of \ac{DHCAL}
performance in terms of single hadron and jet energy resolution, two-particle separation, etc.

\section*{Acknowledgments}
This study is supported by the Minerva Foundation with funding from the Federal German Ministry for Education and Research, as well as by the Krenter-Perinot Center for High-Energy Particle Physics. Additional support comes from a research grant provided by Shimon and Golde Picker, the Nella and Leon Benoziyo Center for High Energy Physics, and the Sir Charles Clore Prize. Special thanks go to Martin Kushner Schnur for his invaluable support of this research.


\begin{thebibliography}{99}
\bibitem{Wigmans} R. Wigmans. Calorimetry. Oxford Science Publications, 2000.
\bibitem{Thomson:2009rp}  M. A.Thomson, Particle Flow Calorimetry and the PandoraPFA Algorithm. NIM \textbf{A 611}, 25--40 (2009).
\bibitem{Dans} D. Shaked-Renous et al., Test-beam and simulation studies towards RPWELL-based DHCAL. JINST \textbf{17}, P12008 (2020).
\bibitem{Fe_calice} CALICE Collaboration, Analysis of Testbeam Data of the Highly Granular RPC-Steel CALICE Digital Hadron Calorimeter and Validation of Geant4 Monte Carlo Models. NIM \textbf{A 939}, 89–105(2019).
\bibitem{geant4} S. Agostinelli et al., GEANT4: A Simulation toolkit. NIM \textbf{A506},  250 (2003). 
\bibitem{zaheer2018deepsets} M. Zaheer et al., Deep Sets. arXiv:1703.06114 (2018).
\bibitem{vaswani2023attentionneed} A. Vaswani et al., Attention Is All You Need. arXiv:1706.03762 (2023).
\bibitem{pytorch} A. Paszke et. al., Automatic differentiation in PyTorch. NIPS 2017 Workshop Autodiff, October 2017.
\bibitem{tensorflow} M. Abadi et al., Tensorflow: A system for large-scale machine learning, doi.org/10.5281/zenodo.5043456, 2016.
\bibitem{softmax} Y. LeCun, L. Bottou, G.B. Orr and K.-R. Müller, Efficient backprop, in Neural networks: Tricks of the trade, pp. 9–50, Springer (2002)
\end{thebibliography}

\end{document}